\documentclass[referee]{aa}

\def\Mesz{M\'esz\'aros~}

\def\beq{\begin{equation}}
\def\enq{\end{equation}}
\def\bea{\begin{eqnarray}}
\def\ena{\end{eqnarray}}
\def\bec{\begin{center}}
\def\enc{\end{center}}
\def\etal{{et al.~}}
\def\eps{\epsilon}
\def\ob{{\rm ob}}

\begin{document}

\title{Long-term Neutrino Afterglows from Gamma-Ray Bursts}

\author{ Zhuo~Li, Z.G.~Dai and T.~Lu }

\offprints{ Zhuo~Li (E-mail: lizhuo@nju.edu.cn) }

\institute{ Department of Astronomy, Nanjing University, Nanjing 210093,
        P. R. China }

\thesaurus{02.01.1, 02.05.1, 13.07.1}

\date{Received  / accepted }

\maketitle \markboth{Z. Li et al.: Neutrino Afterglows from GRBs}{}

\begin{abstract}
It is widely believed that multiwavelength afterglows of gamma-ray bursts (GRBs) originate from relativistic blast
waves. We here show that in such blast waves, a significant fraction of the energy of shock-accelerated protons
would be lost due to pion production by interactions with afterglow photons. This could lead to long-term
production of $10^{16}$--$10^{18}$~eV neutrinos and sub-TeV $\gamma$-rays that accompany with usual afterglows,
provided that the protons are accelerated to $10^{19}$~eV in the blast waves.
\end{abstract}


\section{Introduction}
In the standard fireball shock model, GRBs and their afterglows are produced due to the dissipation of kinetic
energy of relativistic ejecta with Lorentz factor of $\gamma_0 \sim 100-1000$, so-called ``fireball", released
from an unknown central engine (see Piran 1999; Paradijs, Kouveliotou, \& Wijers 2000; \Mesz 2002 for detailed
reviews). In this model, the shocks, resulting from the collisions between shells with different Lorentz factors
in the ejecta (internal shocks) or between the ejecta and the surrounding medium (external shocks), accelerate the
electrons responsible for prompt gamma-rays in GRBs and for X-ray, optical and radio emissions of afterglows. The
same shocks should also accelerate protons. It is pointed out that the physical conditions in the shocks allow
protons to be accelerated to $\eps_p \sim 10^{20}$ eV (Waxman 1995a; Vietri 1995; Dermer \& Humi 2001. See but
Stecker 2000; Mannheim 2000; Scully \& Stecker 2002). Furthermore, the spectrum and flux of ultra-high energy
cosmic rays (UHECRs) are consistent with those expected from Fermi acceleration of protons in GRBs (Waxman 1995b),
suggesting a common origin of GRBs and UHECRs. This picture will lead obviously to high energy neutrinos produced
by $\pi^+$ created in photo-pion interactions between the accelerated protons in the internal or external shocks
and synchrotron photons from electrons accelerated by the same shocks (Waxman \& Bahcall 1997, 2000; Halzen 1998;
Vietri 1998ab; Dai \& Lu 2001; see Waxman 2001 for a review). This $p \gamma$ reaction has a peak at the energy
threshold of the photo-meson $\Delta$-resonance, i.e. $\eps_{\gamma}\eps_{p}=0.2{\rm GeV}^2 \gamma^2$, and the
resulting neutrinos own $\sim 5\%$ of the proton energy in the decay of charged pions and muons. This leads to
neutrinos of $\sim 10^{14}$ eV for prompt gamma-rays from internal shocks (Waxman \& Bahcall 1997; Halzen 1998),
while a UV/optical flash from the external reverse shock implies higher energy neutrinos (e.g., $\sim 10^{18}$ eV)
(Waxman \& Bahcall 2000). Vietri (1998a, 1998b) studied the neutrino production in the case of external forward
shocks, and pointed out that the  neutrino energy may exceed $10^{19}$ eV. But Waxman \& Bahcall (1999) argued
that the contributions of external shocks to the neutrino flux is small.

In recent years, the standard afterglow model, which is significantly successful in interpreting observations, has
been developed (\Mesz \& Rees 1997; Sari, Piran \& Narayan 1998; Wijers \& Galama 1999). In this paper, we
re-investigate the case of external shock neutrinos based on the present detailed afterglow model. We derive the
maximum energy that protons could reach by Fermi acceleration and the efficiency of energy loss from accelerated
protons to pions in the forward shock. By considering the cooling of secondary particles, $\pi^+$ and $\mu^+$, we
derive the spectrum and the flux of the resulting neutrinos. We find that at late times, e.g., $\sim 1$ day after
the GRB trigger, the forward shock may be able to produce an intense neutrino signal, even though the flux is
smaller at earlier times, as stated by Waxman \& Bahcall (1999).

We discuss in section 2 the standard afterglow model, including the dynamics  and synchrotron radiation of the
blast wave, and in section 3 the proton acceleration by the same shock. In section 4 we discuss the neutrino
production rate from the shock-accelerated protons and the maximum energy that neutrinos may achieve. The neutrino
spectrum, flux and detectability are derived in section 5. Conclusions and discussion are given in section 6.

\section{Dynamics and emissions of afterglows}
The fireball shock model suggests that after the prompt burst from internal shocks, the ejecta with total kinetic
energy $E$ continues to expand and drives a relativistic blast wave into the surrounding medium. Once the shocked
medium gets an energy similar to $E$, the blast wave will enter the Blandford-McKee (1976) self-similar evolution,
where the radius $r$ and the Lorentz factor $\gamma$ evolve with observer time as (Wijers \& Galama 1999)
\begin{equation}\label {r}
r=4.07\times 10^{17}(E_{53}/n_0)^{1/4}t_d^{1/4}{\rm cm},
\end{equation}
\begin{equation}\label {gamma}
\gamma=8.87(E_{53}/n_0)^{1/8}t_d^{-3/8},
\end{equation}
where $n$ is the baryon number density of the medium, and $t_d=t/1 ~{\rm day}$. Unless stated otherwise, the
useful form $Q=10^xQ_x$ has been used in this paper.

Let $\xi_e$ and $\xi_B$ be the fractions of the post-shock energy density $U=4\gamma^2 nm_p c^2$ (in the comoving
frame with the shocked material) that are carried by electrons and magnetic fields respectively. The
characteristic electron Lorentz factor (in the comoving frame of the shocked material) is $\gamma_m \simeq \xi_e
\gamma m_p/m_e$, where the post-shock energy per proton is $\gamma m_p c^2$. The strength of the magnetic field in
the comoving frame is
\begin{equation}\label {B}
B=0.34\xi_{B,-2}^{1/2}E_{53}^{1/8}n_0^{3/8}t_d^{-3/8}{\rm G}.
\end{equation}

From the above equations, we can calculate the evolution of synchrotron spectrum from the blast wave. The
characteristic energy of synchrotron photons from electrons with Lorentz factor $\gamma_m$ is
\begin{equation}\label{eps_m}
\eps_m^{\rm ob}\simeq\hbar\gamma\gamma_m^2\frac{eB}{m_ec}
=0.09\xi_{e,-1}^2\xi_{B,-2}^{1/2}E_{53}^{1/2}t_d^{-3/2}{\rm eV},
\end{equation}
where and hereafter we denote the particle energy in the observer frame with superscript ``ob", and particle
energy measured at the comoving frame without superscript. The characteristic energy of synchrotron photons from
electrons with the cooling time, $6\pi m_ec/\sigma_T \gamma_c B^2$, which equals the blast wave expansion time
($\sim r/\gamma c$), is given by
\begin{equation}
\eps_c^{\rm ob}\simeq\hbar\gamma\gamma_c^2\frac{eB}{m_ec} =3\xi_{B,-2}^{-3/2}n_0^{-1}E_{53}^{-1/2}t_d^{-1/2}{\rm
eV},
\end{equation}
and the specific luminosity $L_{\eps}=dL/d\eps^{\rm ob}$ at the peak is
\begin{equation}\label{L_max}
       L_{max} \simeq (2\pi \hbar)^{-1} \gamma \frac {e^3B}{m_ec^2}N_e
          \simeq 5\times 10^{58}(\xi_{B,-2}n_0)^{1/2}E_{53}~{\rm s}^{-1},
\end{equation}
where $N_e={4\over 3} \pi r^3 n$ is the total number of the swept-up electrons. Eq. (\ref{L_max}) implies that the
peak specific luminosity is a time-independent constant. The electrons are accelerated by the shock to a power-law
energy distribution, $dN_e/d\gamma_e\propto \gamma_e^{-p}$ for $\gamma_e
>\gamma_m$, where $p\simeq 2.2$ is fitted to the observations of afterglows. We take $p=2$ for simplicity here.
After $t\sim 1$ hour, $\eps_c^{\rm ob}>\eps_m^{\rm ob}$. In this case, the specific luminosity $L_{\eps}$ peaks at
$\eps_m^{\rm ob}$, $L_{max}=L_m$, and the synchrotron spectrum is a broken power law (Sari, Piran \& Narayan 1998)
as: $L_{\eps}=L_m(\eps/\eps_m)^{-1/2}$ for $\eps_m <\eps <\eps_c$ and $L_{\eps}\propto \eps^{-1}$ for $\eps
>\eps_c$. Note that in the above equations, the typical parameter values have been adopted as
$\xi_e=10^{-1}\xi_{e,-1}$, $\xi_B=10^{-2}\xi_{B,-2}$, $E=10^{53}E_{53}$~erg and $n=1n_0$~cm$^{-3}$, which are
inferred from some afterglow observations (e.g. Wijers \& Galama 1999; Granot, Piran \& Sari 1999).

\section{High-energy proton production}
Following the analysis of Waxman (1995a) in the case of internal shocks, we investigate the physical condition in
external forward shocks for high energy protons production. The typical Fermi acceleration time is $t_a = fR_L/c$,
where the Larmor radius is given by $R_L=\eps_p/eB$ and $f$ is of order unity (Hillas 1984). The protons are
available for acceleration only within a comoving expansion time $t_d \sim r/\gamma c$. This leads to the
requirement that $t_a<t_d$. From Eqs. (\ref{r})-(\ref{B}), we have
\begin{equation}\label{pmax1}
\eps_{p}^{\rm ob}<4\times 10^{19}\xi_{B,-2}^{1/2}E_{53}^{3/8}n_0^{1/8}t_d^{-1/8}{\rm eV}.
\end{equation}
If a magnetic field is nearly in equipartition (i.e., $\xi_B\sim1$), the protons can be accelerated to well above
$10^{20}$~eV. However, fits of afterglow observations give a typical value $\xi_B=0.01$, showing that the magnetic
fields are too weak to accelerate protons up to $10^{20}$~eV in the blast waves. This upper limit is somewhat
robust since its dependence on the medium density and the time ($\propto n^{1/8}t^{-1/8}$) is very weak.

The proton acceleration is affected by energy loss  due to synchrotron radiation. The requirement that the proton
synchrotron cooling time, $t_{p,syn}=(6\pi m_p^4c^3/\sigma_Tm_e^2)\eps _p^{-1}B^{-2}$, should be larger than the
acceleration time $t_a$ is
\begin{equation}\label{pmax2}
\eps_{p}^{\rm ob}<3.2\times 10^{21}\xi_{B,-2}^{-1/4}E_{53}^{1/16}n_0^{-5/16}t_d^{-3/16}{\rm eV}.
\end{equation}
The proton acceleration is also limited by energy loss due to interaction with afterglow photons. In the next
section, we show that, provided a typical medium density $n_0=1$, the fraction of energy loss of a proton in the
expansion time $t_d$ is usually not larger than one, which implies that the limit from energy loss due to
photo-pion interaction is a less stringent constraint than that from $t_a<t_d$. Thus, the maximum energy of
protons comes from Eq. (\ref{pmax1}), $\eps_{p,max}^\ob\simeq4\times10^{19}$~eV.

Gallant \& Achterberg (1999) studied the shock-acceleration, also called first order Fermi-acceleration, in GRBs
and argued that it may be impossible for protons to be shock-accelerated to above $10^{18.5}$ eV in an
ultra-relativistic blast wave which is driven by the fireball ejecta into a typical interstellar medium. However
we still expect protons to be accelerated to ultra-high energy in the blast waves by second order
Fermi-acceleration, where the protons gain energy from the scattering at plasma waves (e.g. Dermer \& Humi 2001).


\section{Neutrino production}
Denoting the photon number density in the comoving frame of the shocked medium by $n_{\gamma}(\eps)d\eps$ and
following Waxman \& Bahcall (1997), we can write the fractional energy-loss rate of a proton with energy $\eps_p$
due to pion production,
\begin{eqnarray}
t_\pi^{-1}(\epsilon_p)\equiv&& -{1\over\epsilon_p}{d\epsilon_p\over dt}\nonumber\\=&&
{1\over2\gamma_p^2}c\int_{\epsilon_0}^\infty d\epsilon\,\sigma_\pi(\epsilon)
\xi(\epsilon)\epsilon\,\int_{\epsilon/2\gamma_p}^\infty dx\, x^{-2}n(x)\,, \label{pirate}
\end{eqnarray}
where $\gamma_p=\epsilon_p/m_pc^2$, $\sigma_\pi(\epsilon)$ is the cross section for pion production for a photon
with energy $\epsilon$ in the proton rest frame, $\xi(\epsilon)$ is the average fraction of energy lost to the
pion, $\epsilon_0=0.15 \, {\rm GeV}$ is the threshold energy, and the proton number density is related to the
observed specific luminosity by $n(x)=L_\epsilon(\gamma x)/(4\pi r^2c\gamma x)$. Because of the
$\Delta$-resonance, the photo-meson production for $\eps_0/2\eps_{c}\ll\gamma_p<\eps_0/2\eps_{m}$ is dominated by
interaction with photons in the energy range $\eps_{m}<\eps\ll\eps_{c}$, where $L_\eps\propto\eps^{-1/2}$. For
this photon spectrum, Eq. (\ref{pirate}) leads to
\begin{equation}
t_\pi^{-1}(\eps_p)\simeq {2^{3/2}\over2.5} {L_m\over4\pi r^2\gamma} \left({\eps_{\rm peak}\over \gamma_p
\eps_{m}}\right)^{-1/2}
 {\sigma_{\rm peak}\xi_{\rm peak}
\Delta\eps\over\eps_{\rm peak}}. \label{pirate2}
\end{equation}
Here, $\sigma_{\rm peak}\simeq 5\times10^{-28}{\rm cm}^2$ and $\xi_{\rm peak}\simeq0.2$ at the resonance
$\epsilon=\epsilon_{\rm peak}=0.3 \, {\rm GeV}$, and $\Delta\epsilon\simeq0.2 \, {\rm GeV}$ is the peak width. The
fraction of energy loss of a proton with observed energy $\eps_p^{\rm ob}$ by pion production, $f_\pi (\eps_p^{\rm
ob})$, is estimated by $t_{\pi}^{-1}(\eps_p)$ times the comoving expansion time of the shocked material ($\sim
r/\gamma c$). Thus, according to Eqs. (\ref{r}), (\ref{gamma}), (\ref{eps_m}) and (\ref{L_max}), we obtain
\begin{equation}\label{f_pi}
f_\pi(\eps_p^{\rm ob})\approx 0.07\xi_{e,-1}\xi_{B,-2}^{3/4}E_{53}^{5/8}n_0^{9/8}t_d^{1/8} (\eps_{p,19}^{\rm
ob})^{1/2}.
\end{equation}
Eq. (\ref{f_pi}) is valid for protons in the energy range
\begin{eqnarray}
6\times10^{17}\xi_{B,-2}^{3/2}E_{53}^{3/4}n_0^{3/4}t_d^{-1/4}{\rm eV}
<\eps_{p}^{\rm ob}\nonumber\\
<5\times10^{19}\xi_{e,-1}^{-2}\xi_{B,-2}^{-1/2}E_{53}^{-1/4}n_0^{-1/4}t_d^{3/4}{\rm eV} \,. \label{ep range}
\end{eqnarray}
At lower energy, protons interact mainly with photons in the energy range $\eps >\eps_c$, where
$L_\eps\propto\eps^{-1}$, and $f_\pi\propto\eps_p^{\rm ob}$. Eqs. (\ref{f_pi}) and (\ref{ep range}) imply that the
fraction of energy loss, $f_{\pi b}$, at the ``break" energy $\eps_{pb}^{\rm ob}=
6\times10^{17}\xi_{B,-2}^{3/2}E_{53}^{3/4}n_0^{3/4}t_d^{-1/4}{\rm eV}$, is a time-independent constant,
\begin{equation}
f_{\pi b}\approx0.02\xi_{e,-1}\xi_{B,-2}^{3/2}E_{53}n_0^{3/2}.
\end{equation}

The maximum energy of the resultant neutrinos is analyzed as follows. This energy is first determined by the
maximum energy that the accelerated protons achieve. From Eq. (\ref{pmax1}), the maximum energy of neutrinos is
$\sim 2\times 10^{18}$ eV. This energy is also limited by the energy loss of secondary particles, i.e., pions and
muons, because both pions and muons may suffer synchrotron and inverse-Compton (IC) losses before decay (Rachen \&
\Mesz 1998).  For synchrotron losses of pions and muons to be negligible, their Lorentz factor (in the comoving
frame of shocked medium) should not exceed the critical Lorentz factor $\gamma_{\star}$ given by
$\gamma_\star^2\tau_\star=\gamma_pt_{p,syn}(m_\star/m_p)^3$ (Rachen \& \Mesz 1998), where $\star$ refers to either
pions or muons, $\tau_\pi=2.6\times10^{-8}$ s and $\tau_\mu=2.2\times10^{-6}$ s are respective lifetimes in their
rest frames, and $m_\star/m_p\simeq 0.1$ for both pions and muons. Due to the longer lifetime compared to pions,
muons suffer greater energy loss, resulting in lower upper limit of neutrino energy. Using Eq. (\ref{B}), we have
(in the observer frame) the critical energy for muons
\begin{equation}
\eps_{\mu,syn}^{\rm ob}=10^{20}\xi_{B,-2}^{-1/2}n_0^{-1/2}~{\rm eV},
\end{equation}
which is independent of the total fireball energy $E$ and observer time $t$. For IC loss to be negligible, the
critical Lorentz factor is given by $\gamma_\star^2\tau_\star=\gamma_pt_{p,syn}(m_\star/m_p)^3(U_B/U_{ph})$, where
$U_B=\xi_BU$ and $U_{ph}=\xi_e\eta U$ are the energy densities of the magnetic field and afterglow photons,
respectively, and $\eta$ is the fraction of the electron energy that is radiated away. Note that
$\eta=(\eps_c/\eps_m)^{(2-p)/2}$ for slow cooling electrons with $\eps_c>\eps_m$. We take $p\approx2.2$. $\eta$ is
not sensitive to other parameters as $\eta\sim1$. (In fact, $\eta=1$ shows the strongest IC loss and provides the
strongest constraint). Thus, the critical Lorentz factor is
$\gamma_\star^{IC}\approx\gamma_\star^{syn}(\xi_B/\xi_e)^{1/2}$, or
\begin{equation}
\eps_{\mu,IC}^{\rm ob}=4\times10^{19}\xi_{e,-1}^{-1/2}n_0^{-1/2}~{\rm eV},
\end{equation}
which is independent of $\xi_B$. The IC loss is somewhat stronger than the synchrotron loss provided
$\xi_B/\xi_e<1$. But, the energy limit from both losses exceeds the maximum proton energy from Eq. (\ref {pmax1}),
so the neutrino production does not suffer the energy losses of secondary particles. The maximum neutrino energy
is determined by the maximum proton energy, $\eps_{\nu,max}^\ob\simeq2\times 10^{18}$ eV.

\section{Neutrino spectrum, flux and detectability}
The photo-meson interactions include (1) production of $\pi$ mesons: $p\gamma\rightarrow p+\pi^0$ and
$p\gamma\rightarrow n+\pi^+$, and (2) decay of $\pi$ mesons: $\pi^0\rightarrow 2\gamma$ and $\pi^+\rightarrow
\mu^++\nu_\mu\rightarrow e^++\nu_e+ \bar{\nu}_\mu+\nu_\mu$. These processes produce neutrinos with energy $\sim
5\%\epsilon_p$ (Waxman \& Bahcall 1997). Eq. (\ref{ep range}) implies that for a fixed time, the spectrum of
neutrinos below $\eps_{\nu b}^{\rm ob}\approx 3\times 10^{16}\xi_{B,-2}^{3/2}E_{53}^{3/4}n_0^{3/4}t_d^{-1/4} {\rm
eV}$ is harder by one power of the energy than the proton spectrum, and by half a power of the energy at higher
energy. Therefore, if the differential spectrum of accelerated protons is a power law form $n(\eps_p)
\propto\eps_p^{-2}$ (Blandford \& Eichler 1987), the differential neutrino spectrum is $n(\epsilon_\nu)\propto
\epsilon_\nu^{-1}$ below the break and $n(\eps_\nu)\propto \eps_\nu^{-3/2}$ above the break. Unlike the prompt
bursts of neutrinos from either internal or reverse shocks, which has a short duration similar to the GRBs, the
neutrino emission from forward shocks last for a longer time due to the long-term forward shocks. Without the
detailed knowledge of the evolution of the spectrum of accelerated protons in the forward shocks, it is far from
calculating clearly the neutrino fluxes. However, as $\eps_{\nu b}^{\rm ob}(\propto t^{-1/4})$ and
$\eps_{p,max}^\ob(\propto t^{-1/8})$ do not change much from one day to even weeks which is concerned about here,
one can expect a time-integrated neutrino spectrum similar to the spectrum at $t=1$~day.  For simplicity, the
particle energies, hereafter in this section, refer to the value when $t=1$~day, e.g. $\eps_{\nu
b}^\ob\rightarrow\eps_{\nu b}^\ob(t_d=1)$, $\eps_{pb}^\ob\rightarrow\eps_{pb}^\ob(t_d=1)$, etc.

We derive the neutrino fluxes as follows. Defining by $\bar f_\pi\equiv\int n(\eps_p)\eps_p f_\pi d\eps_p/\int
n(\eps_p)\eps_p d\eps_p$ the fraction of energy loss of all accelerated protons to neutrinos, and considering the
neutrino spectrum given in the above paragraph, we obtain, after some calculations,
\begin{equation}
\bar f_\pi\simeq2f_{\pi b}\frac{(\eps_{p,max}^\ob/\eps_{pb}^\ob)^{1/2}}{ \ln(\eps_{p,max}^\ob/\eps_{pl})},~~{\rm
for}\,\, \eps_{pl}\leq\eps_{pb}^\ob.
\end{equation}
Here $f_{\pi b}$ is a constant for a certain GRB, therefore this value is mainly bound to $\eps_{pl}$ which is
determined by the lower threshold of a detector. In this equation, $\eps_{pb}^\ob$ should be replaced by
$\eps_{pl}$ for $\eps_{pl}\geq\eps_{pb}^\ob$. Taking $\eps_{pl}=\eps_{pb}^\ob$, we have $\bar f_\pi\approx0.07$.

The present day muon neutrino energy flux due to GRBs is approximately given by $J\approx0.25(c/4\pi)\bar
f_\pi\dot{E}t_H$, where $t_H\approx10$ Gyr is the Hubble time, and $\dot{E}$ is the production rate of UHECRs in
GRBs per unit volume. The factor $0.25$ here accounts for that about one half of the proton energy is lost to
neutral pions which do not produce neutrinos, and about one half of the energy in charged pions is converted to
$\nu_\mu$'s and ${\bar{\nu}}_\mu$'s. We take $\dot{E}\sim 10^{44}\dot{E}_{44}{\rm erg}\, {\rm Mpc}^{-3}\,{\rm
yr}^{-1}$ (Waxman 1995b). Integrating the specific neutrino flux $\Phi_\nu$ over the neutrino spectrum
($\Phi_\nu\propto[\eps_\nu^{\rm ob}]^0$ below $\eps_{\nu b}^{\rm ob}$, and $\Phi_\nu\propto[\eps_\nu^{\rm
ob}]^{-1/2}$ above $\eps_{\nu b}^{\rm ob}$), we have $J=\int\Phi_\nu d\eps_\nu\simeq2(\eps_{\nu b}^\ob
\eps_{\nu,max}^\ob)^{1/2}\Phi_{\nu b}$. Combining these two equations about $J$ one obtains that the neutrino flux
at $\eps_{\nu b}^{\rm ob}$ is
\begin{equation}
\Phi_{\nu b} \approx 2\times 10^{-18}\frac{\bar f_\pi} {0.07}\dot{E}_{44}\left(\frac{\eps_{\nu b}^\ob} {3\times
10^{16}{\rm eV}}\frac{\eps_{\nu,max}^\ob} {2\times 10^{18}{\rm eV}}\right)^{-1/2}{\rm cm}^{-2} \,{\rm
s}^{-1}\,{\rm sr}^{-1}.
\end{equation}
The resulting high-energy neutrinos may be observed by detecting the Cherenkov light emitted by upward moving
muons produced by neutrino interactions below a detector on the surface of the Earth (Gaisser, Halzen \& Stanev
1995; Gandhi et al. 1998). Planned 1 km$^3$ detectors of high energy neutrinos include ICECUBE, ANTARES, NESTOR
(Halzen 1999) and NuBE (Roy, Crawford \& Trattner 1999). The probability that a neutrino could produce a
high-energy muon in the detector is approximated by $P_{\nu\rightarrow \mu}\approx 10^{-3}(\eps_\nu/10^{16}{\rm
eV})^{1/2}$. We obtain the observed neutrino event rate in a detector,
\begin{eqnarray}
N_{\rm events}(>\eps_{\nu b}^\ob)&=2\pi \int\frac{\Phi_\nu}{\eps_\nu^\ob} P_{\nu\rightarrow \mu}d\eps_\nu^\ob
\approx 2\pi\Phi_{\nu
b}P_{\nu\rightarrow \mu}(\eps_{\nu b}^\ob)\ln\left(\frac{\eps_{\nu,max}^\ob}{\eps_{\nu b}^\ob}\right)\nonumber\\
&\approx 0.04 \frac{\bar f_\pi} {0.07}\dot{E}_{44}\left(\frac{\eps_{\nu,max}^\ob}{2\times 10^{18} {\rm
eV}}\right)^{-1/2}{\rm km}^{-2}\,{\rm yr}^{-1}.
\end{eqnarray}

\section{Conclusions and discussion}
After a prompt $\gamma$-rays burst, the fireball ejecta continues to drive a relativistic blast wave into the
surrounding medium. The shock wave accelerates the electrons to give rise to observed afterglow by synchrotron
radiation. The protons in the shock are also expected to be Fermi-accelerated at the same time. However, the
maximum proton energy may not well exceed $\sim 10^{20}$~eV based on the typical parameters of afterglows. The
interaction of the protons with synchrotron photons leads to a significant fraction of proton energy loss to
neutrino production at energy $10^{16}$--$10^{18}$~eV. Though the synchrotron photon energy density in the blast
wave is much lower than in the reverse shock, the efficiency of proton energy loss in the blast wave is still
comparable to that in the reverse shock, because of the long expansion time of the blast wave available for energy
loss of protons. Since both shocks carry comparable energies (Sari \& Piran 1999), we assume the same production
rate of high-energy cosmic rays, $\dot {E}_{44}\simeq1$ for blast waves as Waxman \& Bahcall (2000) for reverse
shocks. Thus, the expected detection rate of muon neutrinos $0.04$~km$^{-2}$~yr$^{-1}$ is also comparable to the
reverse shock. These mean that besides the internal shock and the reverse shock, the blast wave is also one region
suitable for intense neutrino production.

Waxman \& Bahcall (1999) have estimated a much lower efficiency of the blast wave, i.e. $f_\pi\sim10^{-4}$, using
Eq. (4) of Waxman \& Bahcall (1997). However, the parameter values they chose are only valid for the early phase
when the external shock just forms. When a set of parameters, which are valid for afterglows one day after the
burst, are chosen as follows: $L_\gamma=10^{45}{\rm erg/s},~\Delta t=10^5{\rm s},~\eps_{\gamma b}^\ob=0.1{\rm eV~
and}~\Gamma=10$, their equation gives an efficiency $f_\pi\sim10^{-2}$ in general agreement with ours by orders of
magnitude.

It may be that UHECRs can not  be produced in an ultra-relativistic forward shock (Gallant \& Achterberg 1999)
during the transition of the shock to the self-similar expansion. However the hard X-rays from the forward shock
will interact with the protons accelerated by the reverse shock, yielding a new component of neutrino emission
besides those produced by interaction between protons and optical/UV photons from the reverse shock, which is
proposed by Waxman \& Bahcall (2000). At the moment of transition, especially for the photons at energy $\eps^{\rm
ob}\ga\eps_m^{\rm ob}\simeq 80$ keV (cf. Eq. [\ref {eps_m}]), the specific luminosity from the forward shock
dominates that from the reverse shock, which implies a higher neutrino flux at energy
$\eps_\nu^\ob\la10^{15}(\gamma_0/250)(\eps_m^{\rm ob}/100{\rm keV})^{-1}$ eV.

Different models for progenitors of GRBs include different environments: compact object mergers occur in the ISM
and the explosions of massive stars in the preexisting stellar wind or denser medium. Dai \& Lu (2001) have found
that the neutrino production from reverse shocks in the wind case has a quite different spectrum and 10 times more
detection rate compared with the ISM case, which provides a way to distinguish between GRB progenitor models. In
the blast-wave case discussed in this paper, the proton energy loss rate shows a strong dependence on density
$n$(Eq. [\ref{f_pi}]), we expect much higher fluxes for a denser medium. For $n_4= 1$ typically for giant
molecular clouds, following the analysis in this paper, the energy loss efficiency of a proton at one day after
the the GRB trigger is given by $f_\pi\approx 8.6\eps_{p,18}^\ob t_d^{1/4}$ for $\eps_p^\ob<5\times
10^{18}t_d^{3/4}$~eV, implying that protons of energy $\eps_p^\ob>10^{17}$~eV lose all their energy to pion
production and the neutrino spectrum for $\eps_\nu^\ob>5\times 10^{15}$~eV will trace the proton one. The maximum
energy that protons can be accelerated to, which is now confined by the pion production, is
$4\times10^{18}t_d^{-3/4}$~eV. This implies a maximum neutrino energy of $\sim 10^{17}$~eV.

Another consequence of the photo-meson interactions is a sub-TeV afterglow. The decay of neutral pions produced in
the blast waves would lead to $\sim 10^{18}$~eV photons. Such high energy photons would suffer pair production in
the afterglows because the optical depth for photons above $\sim1$~TeV is important (Zhang \& \Mesz 2001). So
these ultra-high-energy photons would be degraded and leak out as a long-term sub-TeV afterglow, carrying away
$\sim10\%$ energy of protons, similar to the energy losses to neutrinos. For a GRB at typical redshift $z=1$, the
intergalactic diffuse radiation field absorbs the photon above $100$~GeV (Mannheim, Hartmann \& Funk 1996),
allowing only a multi-GeV afterglow observed on the earth. An EGRET observation of GeV emission extended to 1.5
hours after the trigger of GRB 940217 (Hurley \etal 1994) has indicated that a high energy spectral component can
extend to such high energy band and persist for a long period of time. However, the electron-IC emission in the
blast wave can also contribute to this GeV observation. For the parameters,
$n_0\simeq\xi_{e,-1}\simeq\xi_{B,-2}\simeq1$, adopted in this paper, the electron-IC emission is important and
plausible to explain the observation (Zhang \& \Mesz 2001). From the afterglow model presented in section 2, we
can see that the peak energy in the IC emission spectrum, $\gamma_c^2\eps_c^\ob$ (Sari \& Esin 2001), is still
marginally up to 100 GeV even one day after the GRB trigger, but we do not expect GeV IC emission a few days
later, while the emission produced by neutral pion decay might still be possible. In principle, the GeV emission
could also be due to the proton synchrotron emission, but it is not important for the parameters adopted here
(Zhang \& \Mesz 2001).

\begin{acknowledgements}
The authors thank the referee for very good comments to improve this paper. This work was supported by the
National Natural Science Foundation of China (grants 19825109 and 19973003) and the National 973 Project (NKBRSF
G19990754).
\end{acknowledgements}

\end{document}